%% file: sirocco20poa.tex
\documentclass[runningheads]{llncs}

\usepackage{amsmath}
\usepackage{amssymb}
\usepackage{paralist}
\usepackage[normalem]{ulem}
\usepackage{pifont}
\usepackage{tikz}
\usepackage{makecell}
\usepackage[framemethod=tikz]{mdframed}
\usepackage{subcaption}
\usepackage[noadjust]{cite}
\usepackage[title]{appendix}
\usepackage{comment}

\usetikzlibrary{spline}

\input{comment-macros}
\begin{document}

%
\title{The Value of Information in Selfish Routing}

\author{Simon Scherrer\inst{1} \and Adrian Perrig\inst{1} \and Stefan Schmid\inst{2}}
\institute{Department of Computer Science, ETH Zurich \and Faculty of Computer Science, University of Vienna}

\maketitle

%
\begin{abstract}
	Path selection by selfish agents has traditionally been
	studied by comparing social optima and equilibria in the Wardrop model, i.e., by
	investigating the Price of Anarchy in selfish routing. 
	In this work, we refine and extend the traditional selfish-routing model in order
	to answer questions that arise in emerging
	path-aware Internet architectures. The model enables us to characterize the impact of different degrees of congestion information that users possess. Furthermore, it allows us to analytically quantify the impact of selfish routing, not only on users, but also on network operators. Based on our model, we show that the cost of selfish routing depends on the network topology, the perspective (users versus network operators), and the information that users have. Surprisingly, we show 
	analytically and empirically that less information tends to lower the Price of Anarchy, almost to the optimum. 
	Our results hence suggest that selfish routing has modest social cost even without the dissemination of path-load information.
	
\end{abstract}

\section{Introduction}
\label{sec:introduction}

If selfish agents are free to select communication paths in a network,
their interactions can produce sub-optimal traffic allocations.
A long line of research relating to \emph{selfish routing}~\cite{roughgarden2002bad, roughgarden2003price, roughgarden2007routing}  
has quantified many effects of distributed, uncoordinated path selection by selfish individuals
in networks.
While seminal work on such game-theoretic analyses 
dates back to 
Wardrop~\cite{wardrop1952some},
especially the notion of \emph{Price of Anarchy}, coined by
Koutsoupias and Papadmitriou~\cite{koutsoupias1999worst},
has received much
attention:
the Price of Anarchy
compares the worst possible outcome of individual
decision making, i.e., the worst Nash equilibrium,
to the global optimum, by taking the corresponding cost ratio.
The Price of Anarchy in network path selection is typically measured in terms of
\textit{latency}. 

In this paper, we revisit these concepts to investigate two key
aspects which have been less explored in the literature so far
and are highly relevant for newly emerging path-aware network architectures~(cf. \S\ref{sec:introduction:practical}):

\begin{itemize}
	\item \textbf{\emph{Impact of information:}}
	A fundamental design question of network
	architectures concerns which information about the network state
	should be shared with end-hosts,
	beyond the latency information that can be observed by the end-hosts directly.
	\item \textbf{\emph{Impact on network operators:}}
	While game-theoretic analyses usually revolve around the 
	cost experienced by users, it is also important to understand the impact of
	selfish routing on the network operators' cost.
\end{itemize}

\subsection{Practical Motivation}
\label{sec:introduction:practical}

The traditional question studied in the selfish-routing literature,
namely the efficiency of uncoordinated path selection by selfish agents,
has recently received new relevance in the context of emerging
Internet architectures relying on source-based path selection~\cite{andersen2001resilient,raghavan2004system,xu2006miro,filsfils2015segment}. In particular, the already deployed SCION architecture~\cite{barrera2017scion,PeSzReCh2017} offers extensive path-selection control to users.

Today's Internet infrastructure is based on a forwarding mechanism that grants almost 
exclusive control to the network and almost no control to users
(or \emph{end-hosts}).
In fact, all communication from a given end-host to another end-host
takes place over the \emph{single} AS-level path that BGP (Border Gateway Protocol)
converged on. In the upcoming paradigm of 
\emph{source-based path selection}~\cite{8345560}, network operators supply end-hosts with a pre-selected set of paths to a destination, enabling end-hosts to select a
forwarding path themselves.



Source-based path selection allows end-hosts to select paths
with superior performance to the BGP-generated path~\cite{savage1999end,sosr16,gupta2015sdx}
or to quickly switch to an alternative path upon link failures.
However, a widely shared concern about source-based path selection regards 
the loss of control by network operators, which fear that 
the traffic distribution resulting from individual user decisions might impose considerable cost on both themselves and their customers. 
Another concern is that end-hosts require path-load information in order to perform
path selection effectively, necessitating complex systems for the dissemination of network-state information.
We refine and extend concepts from the selfish-routing literature to 
investigate the validity of these concerns.


\subsection{Our Contributions}

We present a game-theoretic model (\S\ref{sec:model}) which allows us to quantify
not only the Price of Anarchy experienced by end-hosts,
but also to account for the network operators. 
Furthermore, we use our model to explore how end-host information about the network state
affects the Price of Anarchy. 

We find that different levels of information indeed
lead to different Nash equilibria and thus also to different Prices of Anarchy.
Intriguingly, we find that while more information can improve the efficiency of selfish routing in networks with few end-hosts (\S\ref{sec:benefits}), more information tends to induce a \emph{higher} Price of Anarchy in more general settings (\S\ref{sec:harms}). 
Indeed, near-optimal outcomes are typically achieved if end-hosts select paths
based on simple latency measurements of different paths.
These theoretical results suggest that source-based path selection cannot only achieve a good network performance in selfish contexts, but can be realized in a fairly light-weight manner, avoiding the need to distribute much information about the network state. This insight is validated with a case study
on the Abilene topology (\S\ref{sec:case-study}).





\section{Model and First Insights}
\label{sec:model}


\subsection{Model}
\label{sec:model:network}

As in previous work on selfish routing~\cite{roughgarden2002bad, fischer2009adaptive}, our model is inspired by the classic Wardrop model \cite{wardrop1952some}. In this model, the network is abstracted as a graph $G = (A, L)$, where the edges $\ell \in L$ between the nodes $A_i \in A$ represent links. Every link $\ell \in L$ is described by a link-cost function $c_{\ell}(f_{\ell})$, where $f_{\ell}$ is the amount of load on link $\ell$, i.e., a \textit{link flow}.  Typically, link-cost functions are seen as describing the latency behavior of a link. To reflect queuing dynamics, link cost functions are convex and non-decreasing. For every node pair $(A_i, A_j)$, there is a set of paths $P(A_i, A_j)$ that contains all non-circular paths between $A_i$ and $A_j$. Between any node pair $(A_i, A_j)$, there is a demand $d$ shared by infinitely many agents, where each agent is controlling an infinitesimal share of traffic.

However, the traditional Wardrop model is not suitable to analyze traffic dynamics in an Internet context. We thus adapt the Wardrop model into a more realistic model as follows. First, we introduce the concept of \textit{ASes} and \textit{end-hosts}, which allows us to perform a more fine-grained analysis of traffic in an inter-domain network. An AS $A_i \in A$ is represented by a node in the network graph $G$. The AS contains a set of end-hosts, which are the players in the path-selection game. Differently than in the Wardrop model, we allow for non-negligible, heterogeneous demand between end-host pairs in order to accommodate the variance of demand in the Internet. For example in origin-destination pair $\mathit{od} = (e_s, e_t) \in \mathit{OD}$ (short: $(s,t)$), an end-host $e_s \in A_i$ can have a demand $d_{s,t} \geq 0$ towards another end-host $e_t \in A_j$. We also deviate from the Wardrop model by considering a multi-path setting, where the demand $d_{s,t}$ of one agent can be arbitrarily distributed over all paths $p \in P(A_i, A_j)$. The amount of flow from end-host $e_s$ to end-host $e_t$ on path $p \in P(A_i, A_j)$ is denoted as a \textit{path flow} $F_{(s,t),p}$, which must be non-negative, with $\sum_{p \in P(A_i, A_j)} F_{(s,t),p} = d_{s,t}$. 
The set $\Pi(e_s,e_t) \subseteq \Pi$ contains all end-host paths of the form $\pi = \big[(s,t),p\big]$,
where $e_s$, $e_t$ are end-hosts connected by the AS-level path $p$.
All path flows $F_{(s,t),p}$ for an origin-destination pair $(e_s,e_t)$ are collected in a path-flow vector $\mathbf{F}_{s,t} \in \mathbb{R}^{|\Pi(e_s,e_t)|}$. All such path-flow vectors $\mathbf{F}_{s,t}$ are collected in the global \textit{path-flow pattern}  $\mathbf{F} \in \mathbb{R}^{|\Pi|}$. 
A link flow $f_{\ell}$ for link $\ell \in L$ is the sum of the path flows in $\mathbf{F}$ that refer to end-host paths $\pi$ containing link $\ell$, i.e., $f_{\ell} = \sum_{\pi \in \Pi: \ell \in \pi} F_{\pi}$. 

The cost of an end-host path $C_{\pi}$ given a certain path-flow pattern $\mathbf{F}$ is the sum of the cost of all links in the path: $C_{\pi}(\mathbf{F}) = \sum_{\ell \in \pi} c_{\ell}(f_{\ell})$. The cost to end-hosts $C^{\ast}(\mathbf{F})$ from a path-flow pattern $\mathbf{F}$ is the latency experienced by all end-hosts on all the paths to all of their destinations, weighted by the amount of traffic that goes over a given path. This term can be simplified as follows: \vspace{-10pt}

\begin{equation*}
	\begin{split}
		C^{\ast}(\mathbf{F}) &= \sum_{(s,t) \in \mathit{OD}} \sum_{\pi \in \Pi(s,t)} F_{\pi} \cdot C_{\pi}(\mathbf{F})
		= \sum_{\pi \in \Pi} F_{\pi}  \cdot \sum_{\ell\in\pi} c_{\ell}(f_{\ell}) =  \sum_{\ell \in L} f_{\ell} \cdot c_{\ell}(f_{\ell})
	\end{split} 
\end{equation*}

Existing work on selfish routing \cite{roughgarden2002bad, roughgarden2007routing} usually defines total cost in the above sense. However, when analyzing source-based path selection architectures, the network-operator perspective on social cost
is essential. Therefore, we also introduce a social cost function relating to the perspective of network operators. 

The basic idea of the network-operator cost function $C^{\#}$ is to treat links as investment assets.  Thus, the business performance of a link $\ell$ is given by a function $p^{\#}_{\ell}(f_{\ell}) = b^{\#}_{\ell}(f_{\ell}) - c^{\#}_{\ell}(f_{\ell})$, where $b^{\#}_{\ell}$ and $c^{\#}_{\ell}$ are the benefits and costs of a link, respectively. As we investigate effects on the aggregate of network operators, we model the network-operator cost function as follows: $$C^{\#}(\mathbf{F}) = \sum_{\ell \in L} -p^{\#}_{\ell}(f_{\ell}) = \sum_{\ell \in L} c^{\#}_{\ell}(f_{\ell}) - b^{\#}_{\ell}(f_{\ell}) = \sum_{\ell \in L} c_{\ell}(f_{\ell})$$ We justify this formulation as follows. Concerning link costs $c_{\ell}^{\#}$, a central insight is that network-operator costs mostly stem from heavily used links. In volume-based interconnection agreements, excessive usage of a link induces high charges, whereas in peering agreements, excessive usage violates the agreement and triggers expensive renegotiation. Moreover, heavy usage necessitates expensive capacity upgrades. As the latency function $c_{\ell}(f_{\ell})$ indicates the congestion level on link $\ell$, we approximate $c_{\ell}^{\#} \approx c_{\ell}$. The link benefit~$b^{\#}_{\ell}$ captures the link revenue, both revenue from customer ASes and customer end-hosts. In the aggregate, the monetary transfers between ASes (charges paid and received) sum up to zero. Given a fixed market size, the revenue from end-hosts sums up to a constant in the aggregate. Hence, the global benefit $\sum_{\ell \in L} b_{\ell}^{\#}$ is constant and can be dropped, as the absolute level of the network-operator cost is irrelevant for our purposes. 
This convex formulation of $C^{\#}$ allows theoretical analysis.



\subsection{Social Optima}
\label{sec:model:optima}

According to Wardrop~\cite{wardrop1952some, dafermos1969traffic}, 
a socially optimal traffic distribution is reached iff the total cost cannot be reduced by moving traffic from one path to another. In the optimum,
the cost increase on an additionally loaded path at least outweighs the cost reduction from a relieved path. Because the cost functions are convex and non-decreasing, it suffices that this condition holds for an infinitesimal traffic share. 
Adding an infinitesimal amount to the argument of a cost function 
imposes a \textit{marginal cost}, given by the derivative of the cost function. 
A socially optimal traffic distribution is thus reached iff the marginal cost of every alternative path is not smaller than the marginal cost of the currently used paths~\cite{dafermos1969traffic}:

\vspace{1mm}

\begin{mdframed}[hidealllines=true,backgroundcolor=gray!20]
\textbf{Social optimum.} A path-flow pattern $\mathbf{F}$ represents a social optimum w.r.t. cost function $C$ if and only if for every origin-destination pair $\mathit{od} \in \mathit{OD}$, the paths $\pi_1, ..., \pi_i, \pi_{i+1}, ..., \pi_{|\Pi(\mathit{od})|} \in \Pi(od)$ stand in the following relationship:
\begin{equation*}
	\begin{split}
	\frac{\partial}{\partial F_{\pi_1}} C(\mathbf{F}) = ... =  \frac{\partial}{\partial F_{\pi_i}} C(\mathbf{F}) &\leq \frac{\partial}{\partial F_{\pi_{i+1}}} C(\mathbf{F}) \leq ... \leq \frac{\partial}{\partial F_{\pi_{|\Pi(\mathit{od})|}}} C(\mathbf{F})\\
	F_{\pi} > 0 \quad \text{for} \quad \pi = \pi_1, ..., \pi_i, \hspace{25pt}&\hspace{25pt}
	F_{\pi} = 0 \quad \text{for} \quad \pi = \pi_{i+1}, ..., \pi_{|\Pi(\mathit{od})|}.
	\end{split}
\end{equation*}
\end{mdframed}

\vspace{-5pt}
In this work, we refine the conventional notion of the social optimum by distinguishing two different perspectives on social cost: The \textit{end-host optimum}~$\mathbf{F}^{\ast}$ satisfies the above conditions with respect to the function $C^{\ast}$, whereas the \textit{network-operator optimum} $\mathbf{F}^{\#}$ satisfies the above conditions with respect to function $C^{\#}$.\vspace{-15pt}

\begin{figure}
	\centering
	\input{fig/model-example}
	\caption{Example network illustrating the source-based path selection model.}
	\label{fig:introduction:model-example}
	\vspace{-15pt}
\end{figure}
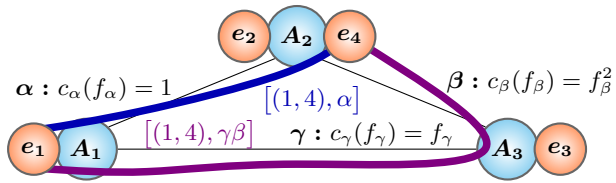

Interestingly, the end-host optimum $\mathbf{F}^{\ast}$ and the network-operator optimum~$\mathbf{F}^{\#}$
can differ substantially. Assume that end-host $e_1$ in Figure~\ref{fig:introduction:model-example} has a demand of $d_{1,4} = 1$ towards end-host $e_4$ and that there is no other traffic in the network. The network-operator cost function $C^{\#}(\mathbf{F})$ is $1 + F_{\gamma\beta}^2 + F_{\gamma\beta}$ and is minimized by $\mathbf{F}^{\#} = (1,0)^{\top}$, i.e., by sending all traffic over link $\alpha$. In contrast, the end-host cost function is $F_{\alpha} + F_{\gamma\beta}^3 + F_{\gamma\beta}^2$ and is minimized by $\mathbf{F}^{\ast} = (2/3,1/3)^{\top}$, i.e., by sending two thirds of traffic over link $\alpha$ and the remaining third over path $\gamma\beta$.

\subsection{Degrees of Information}
\label{sec:model:degrees-information}

In this paper, we consider the following 
two assumptions on the network information 
possessed by end-hosts: 

\begin{itemize}
	\item\textbf{Latency-only information (LI):} End-hosts know the latency of every path to a destination.
	\item \textbf{Perfect information (PI):} End-hosts know not only the latency of different paths, but also how the latency of the network links depends on the current
	load, i.e., the \emph{latency functions}. Moreover, the end-hosts know the current link utilization, i.e., the background traffic.
\end{itemize}

The LI assumption hence reflects a scenario where end-hosts have to rely solely on latency measurements of paths, i.e., 
through RTT measurements from their own device.
 The LI assumption is the standard model traditionally 
considered in the selfish routing literature~\cite{koutsoupias1999worst, roughgarden2002bad, friedman2004genericity}. 

In this work, we extend the standard model by introducing the concept of \emph{perfect information} (PI). The PI assumption reflects a scenario where end-hosts can always take the \textit{best} traffic-allocation decision in selfish terms.
More specifically, the PI assumption allows end-hosts to compute the \textit{marginal cost} of a path. 
In path-aware networking, supplying end-hosts with perfect information is possible, as such information is known by network operators and can be disseminated along with path information.

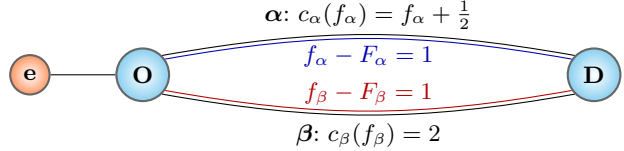
\begin{figure}
	\vspace{-15pt}
	\centering
	\input{fig/introduction-example}
	\vspace{-10pt}
	\caption{Example illustrating the different degrees of end-host information.}
	\label{fig:introduction:information-example}
	\vspace{-15pt}
\end{figure}

Figure~\ref{fig:introduction:information-example} illustrates the difference between the LI assumption and the PI assumption. Assume that end-host $e$, residing in AS $A$, has a demand of $d = 1$ to a destination in AS $D$. End-host $e$ can split its traffic between two paths $\alpha$ and $\beta$, both consisting of a single link with the cost functions $c_{\alpha}$ (linear) and $c_{\beta}$ (constant). The background traffic (traffic not from end-host $e$) is 1 on both paths. Assuming the traffic allocation of end-host $e$ is $(F_{\alpha}, F_{\beta}) = (0.5, 0.5)$, the path-latency values are given by $c_{\alpha}(0.5 + 1) = 2$ and $c_{\beta}(0.5+1) =  2$. Given the LI assumption, end-host~$e$ performs no traffic reallocation, as there is no lower-cost alternative path which traffic could be shifted to. Moreover, there is no method for predicting the path costs for a different traffic allocation. However, such a prediction is possible with perfect information (PI): under the PI assumption, end-host $e$ knows the cost functions and the background traffic such that it can optimize the objective $F_{\alpha} \cdot (F_{\alpha}+1+\frac{1}{2}) + (1 - F_{\alpha}) \cdot 2$. As a result, end-host $e$ discovers the optimal traffic assignment $(0.25, 0.75)$. Intriguingly,  the more detailed perfect information (PI) enables end-host $e$ to detect an optimization that it cannot directly observe when confronted with latency values only (LI). 


\subsection{Nash Equilibria}
\label{sec:model:equilibria}

In general, uncoordinated actions of selfish end-hosts do not result
in socially optimal traffic allocations. Instead, the only stable states that arise in selfish path selection are \textit{Nash equilibria}, i.e., situations in which no end-host perceives an opportunity to reduce its selfish cost by unilaterally reallocating traffic. However, as shown in \S\ref{sec:model:degrees-information}, the degree of available information (LI or PI) strongly influences the optimization opportunities that an end-host perceives. Therefore, different information assumptions induce different types of Nash equlibria:


\vspace{1mm}

\textbf{\textit{LI equilibrium.}} An end-host restricted to latency measurements will shift traffic from high-cost paths to low-cost paths whenever there is a cost discrepancy between paths, and will stop reallocating traffic whenever there is no lower-cost path anymore which the traffic could be shifted to. In the latter situation, an end-host under the LI assumption cannot perceive any way of reducing its selfish cost. A  Nash equilibrium under the LI assumption (short: LI equilibrium) can thus be defined as follows:

\vspace{1mm}

\begin{mdframed}[hidealllines=true,backgroundcolor=gray!20]
\textbf{LI equilibrium.} A path-flow pattern $\mathbf{F}$ represents an LI equilibrium $\mathbf{F}^{0}$ if and only if for every origin-destination pair $\mathit{od} \in \mathit{OD}$, the paths\\$\pi_1, ..., \pi_i, \pi_{i+1}, ..., \pi_{|\Pi(\mathit{od})|} \in \Pi(\mathit{od})$ have the following relationship:
\begin{equation*}
\begin{split}
C_{\pi_1}(\mathbf{F}) = ... =  C_{\pi_i}(\mathbf{F}) &\leq C_{\pi_{i+1}}(\mathbf{F}) \leq ... \leq C_{\pi_{|\Pi(\mathit{od})|}}(\mathbf{F})\\
F_{\pi} > 0 \quad \text{for} \quad \pi = \pi_1, ..., \pi_i \hspace{25pt}&\hspace{25pt}
F_{\pi} = 0 \quad \text{for} \quad \pi = \pi_{i+1}, ..., \pi_{|\Pi(\mathit{od})|}
\end{split}
\end{equation*}
\end{mdframed}

Traditionally, selfish-routing literature~\cite{roughgarden2002bad, roughgarden2003price, friedman2004genericity} considers a Nash equilibrium in the sense of the LI equilibrium, namely an equilibrium defined by the cost equality of all used paths to a destination. Under this classical definition, selfish routing is an instance of a \textit{potential game}~\cite{sandholm2001potential}. 

\vspace{1mm}

\textbf{\textit{PI Equilibrium.}} We contrast the classical equilibrium (LI equilibrium) with a different equilibrium definition that builds on our new concept of perfect information (PI). As explained in \S\ref{sec:model:degrees-information}, the PI assumption states that end-hosts do not only possess cost information of available paths to a destination, but are informed about the cost \textit{functions} of all links in the available paths, as well as the background traffic on these links, i.e., the arguments to the cost functions. An end-host can thus calculate the selfish cost of a specific traffic reallocation and find the path-flow pattern that minimizes the end-host's selfish cost.

The selfish cost $C^{\ast}_{(e)}(\mathbf{F})$ of end-host $e$ is given by the cost of all paths to all desired destinations, weighted by the amount of flow relevant to end-host~$e$:
$$C^{\ast}_{(e)}(\mathbf{F})= \sum_{\ell \in L} f_{\ell,(e)}\cdot c_{\ell}(f_{\ell})$$ where $f_{\ell,(e)}$ is the flow volume on link $\ell$ for which $e$ is origin or destination.

Similar to the end-host social cost function $C^{\ast}$ of which it is a partial term, 
$C^{\ast}_{(e)}$ has a minimum that is characterized by a marginal-cost equality. An equilibrium under the PI assumption is thus given if and only if all end-hosts are at the minimum of their respective selfish cost functions, given the traffic by all other end-hosts:

\vspace{1mm}

\begin{mdframed}[hidealllines=true,backgroundcolor=gray!20]
\textbf{PI equilibrium.}  A path-flow pattern $\mathbf{F}$ represents a PI equilibrium $\mathbf{F}^{+}$ if and only if for every origin-destination pair $\mathit{od} = (e, \_) \in \mathit{OD}$, the paths $\pi_1, ..., \pi_i, \pi_{i+1}, ..., \pi_{P} \in \Pi(\mathit{od})$ stand in the following relationship:
\begin{equation*}
\begin{split}
\frac{\partial}{\partial F_{\pi_1}} C^{\ast}_{(e)}(\mathbf{F}) = ... =  \frac{\partial}{\partial F_{\pi_i}} C^{\ast}_{(e)}\textbf{} &\leq \frac{\partial}{\partial F_{\pi_{i+1}}} C^{\ast}_{(e)}(\mathbf{F}) \leq ... \leq \frac{\partial}{\partial F_{\pi_{|\Pi(\mathit{od})|}}} C^{\ast}_{(e)}(\mathbf{F})\\
F_{\pi} > 0 \quad \text{for} \quad \pi = \pi_1, ..., \pi_i \hspace{25pt}&\hspace{25pt} F_{\pi} = 0 \quad \text{for} \quad \pi = \pi_{i+1}, ..., \pi_{|\Pi(\mathit{od})|}
\end{split}
\end{equation*}
\end{mdframed}

\subsection{Price of Anarchy}
\label{sec:model:poa}

A natural way of analyzing the efficiency of selfish routing is to compare the social optima and the equilibria in a network. Typically, such a comparison involves computing the \textit{Price of Anarchy (PoA)}, i.e., the ratio of the equilibrium cost and the optimal cost. By definition of the optimal cost, this ratio is always larger or equal to~1.

In our model, the classical Price of Anarchy from the existing literature reflects a comparison of the end-host cost $C^{\ast}$ of the LI equilibrium $\mathbf{F}^0$ and the end-host cost $C^{\ast}$ of the end-host optimum $\mathbf{F}^{\ast}$. With the additional versions of social optima and equilibria established in the preceding sections, a total of four different variants of the Price of Anarchy are possible, one for each combination of equilibrium (LI or PI) and perspective (end-hosts or network operators):

\begin{table}
	\vspace{-10pt}
	\renewcommand{\arraystretch}{1.5}
	\centering
	\begin{tabular}{| c | c | c |}
		\cline{2-3}
		 \multicolumn{1}{c|}{} & \textbf{LI equilibrium} & \textbf{PI equilibrium}\\
		\hline
		\makecell{\textbf{End-host perspective}} & \(\mathit{PoA}^{\ast 0} = \frac{C^{\ast}(\mathbf{F}^0)}{C^{\ast}(\mathbf{F}^{\ast})}\)  & \(\mathit{PoA}^{\ast +} = \frac{C^{\ast}(\mathbf{F}^+)}{C^{\ast}(\mathbf{F}^{\ast})}\)\\
		\hline
		\makecell{\textbf{Network-operator perspective}} & \(\mathit{PoA}^{\# 0} = \frac{C^{\#}(\mathbf{F}^0)}{C^{\#}(\mathbf{F}^{\#})}\) & \(\mathit{PoA}^{\# +} = \frac{C^{\#}(\mathbf{F}^+)}{C^{\#}(\mathbf{F}^{\#})}\)\\
		\hline
	\end{tabular}
\vspace{7pt}
	\caption{Different versions of the Price of Anarchy.}
	\label{tab:model:poas}
	\vspace{-15pt}
\end{table}

\subsection{Value of Information}
\label{sec:model:voi}

To compare different equilibria for different information assumptions, we introduce the \textit{Value of Information (VoI)}. For a given perspective, the Value of Information is the difference between the Prices of Anarchy under the LI and PI assumptions, denominated by the Price of Anarchy under the LI assumption:

\begin{equation*}
	\mathit{VoI}^{\ast} = \frac{\mathit{PoA}^{\ast 0}-\mathit{PoA}^{\ast +}}{\mathit{PoA}^{\ast 0}} \quad \quad \mathit{VoI}^{\#} = \frac{\mathit{PoA}^{\# 0}-\mathit{PoA}^{\# +}}{\mathit{PoA}^{\# 0}}
\end{equation*}

A positive Value of Information reflects a situation where the equilibrium under the PI assumption is closer to the social optimum than the equilibrium under the LI assumption. We identify and analyze scenarios with a positive impact of information in \S\ref{sec:benefits}. A negative Value of Information reflects the counter-intuitive scenario where additional information makes the equilibrium more costly (cf.~\S\ref{sec:harms}). 

\section{The Benefits of Information}
\label{sec:benefits}

In this section, we will show that
information is beneficial in the artificial network settings traditionally
considered in the literature~\cite{roughgarden2003price}. 
More precisely, we show that in this setting, the PI equilibrium induces a lower Price of Anarchy than the LI equilibrium such that the Value of Information is positive.
This is intuitive: if end-hosts possess more information, source-based path selection is more efficient.

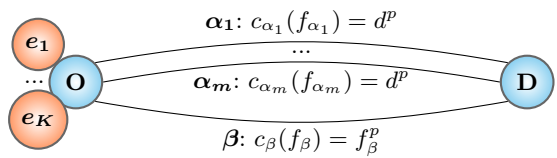
\begin{figure}
	\vspace{-15pt}
	\centering
	\input{fig/network-parallel-links}
	\vspace{-10pt}
	\caption{Example network with beneficial impact of end-host information.}
	\label{fig:benefits:parallel-links}
	\vspace{-10pt}
\end{figure}

In the network of Figure~\ref{fig:benefits:parallel-links}, $K$ end-hosts $e_1$, ...,~$e_K$ reside in AS $O$. Each end-host has a demand of $d/K$ towards a destination in AS $D$. ASes~$O$ and~$D$ are connected by $m$ links $\alpha_1$, ..., $\alpha_m$ with a constant cost function $c_{\alpha_i}(f_{\alpha_i}) = d^p$ and one link $\beta$ with a load-dependent cost function $c_{\beta}(f_{\beta}) = f_{\beta}^p$, where $p \geq 1$. 

Such networks of parallel links are of special importance in the theoretical selfish-routing literature. In particular, Roughgarden~\cite{roughgarden2003price} proved that the network in Figure~\ref{fig:benefits:parallel-links} reveals the worst-case Price of Anarchy for any network
with link cost functions limited to polynomials of degree $p$. The intuition behind this result is that the Price of Anarchy relates to a difference of steepness between cost functions of competing links: the link $\beta$ allows to reduce the cost of traffic from AS $O$ to AS $D$ if used modestly, but loses its advantage over the links~$\alpha_i$ if fully used. However, in selfish routing, end-hosts will use link~$\beta$ until the link is fully used, as it is always a lower-cost alternative path if not fully used. Therefore, the end-hosts overuse link~$\beta$ compared to the optimum. Intuitively, the parallel-links network represents a network where end-hosts have a choice between paths with different latency behavior.

Roughgarden's result refers to the classical Price of Anarchy, i.e., the Price of Anarchy $\mathit{PoA}^{\ast0}$ to end-hosts under the LI assumption. In this section, we will show how this result is affected by additionally introducing the network-operator perspective and the PI assumption. In particular, we will prove the following theorem: \begin{theorem}
	In a network of parallel links, a higher degree of information (PI assumption) is always more socially beneficial compared to a lower degree of information (LI assumption), both from the perspective of end-hosts and network operators: $$\mathit{PoA}^{\ast+} \leq \mathit{PoA}^{\ast0} \hspace{20pt} \mathit{PoA}^{\#+} \leq \mathit{PoA}^{\#0}$$
	\label{theorem:beneficial}
	\vspace{-5pt}
\end{theorem}

\subsection{Social Optima}
\label{sec:benefits:optima}

The end-host optimum in the network of parallel links can be shown to have
social cost $C^{*}(\mathbf{F}^{*}) = d^{p+1}[1 - p/(p+1)^{(p+1)/p}]$. As the derivation is relatively similar to Roughgarden~\cite{roughgarden2003price}, it has been moved to Appendix \ref{app:benefits:endhost-optimum}.

The network-operator optimum $\mathbf{F}^{\#}$ is simple to 
derive: Since the cost of the links $\alpha_i$ is independent of the flow on these links in contrast to the cost of link $\beta$, any flow on link~$\beta$ increases the cost $C^{\#}$ to network operators. The minimal cost to network operators is thus simply $C^{\#}(\mathbf{F}^{\#}) = m \cdot d^p$. 

\subsection{LI Equilibrium}
\label{sec:benefits:li}

Under the LI assumption, a network is in equilibrium if for every end-host pair, all used paths have the same cost and all unused paths do not have a lower cost. Applied to the simple network in Figure~\ref{fig:benefits:parallel-links}, this condition is satisfied if and only if $f_{\beta}^0 = d$ and $f_{\alpha_i}^0 = 0\ \forall f_{\alpha_i}$, implying $c_{\beta}(f_{\beta}^0) = d^p = c_{\alpha_i}(f_{\alpha_i}^0)$. The path-flow pattern $\mathbf{F}^{0}$ with $F_{(k,D),\beta} = d/K$ and $F_{(k,D),\alpha_i} = 0$ represents the LI equilibrium. The cost $C^{\ast}$ of the LI equilibrium $\mathbf{F}^{0}$ to end-hosts is simply $C^{\ast}(\mathbf{F}^0) = d^{p+1}$. The Price of Anarchy to end-hosts under the LI assumption is thus $\mathit{PoA}^{\ast 0} = C^{\ast}(\mathbf{F}^0)/C^{\ast}(\mathbf{F}^{\ast}) = [1 - p/(p+1)^{(p+1)/p}]^{-1}$.

The cost $C^{\#}$ of the LI equilibrium $\mathbf{F}^{0}$ to network operators is given by $C^{\#}(\mathbf{F}^0) =  d^p + \sum_{\alpha_i} d^p = (m+1) \cdot d^p$. The Price of Anarchy to network operators under the LI assumption is thus $\mathit{PoA}^{\# 0} = C^{\#}(\mathbf{F}^0)/C^{\#}(\mathbf{F}^{\#}) = (m+1)/m$, which is maximal for the number $m = 1$ of links $\alpha_i$. The Price of Anarchy to network operators in networks of parallel links is thus upper-bounded by $\mathit{PoA}^{\# 0}_{m = 1} = 2$ whereas the Price of Anarchy to end-hosts is unbounded for arbitrary~$p$. 

\subsection{PI Equilibrium}
\label{sec:benefits:pi}

If the end-hosts $e_1$,..., $e_K$ are equipped with perfect information, they are in equilibrium if and only if the \textit{selfish} marginal cost of every path to AS $D$ is the same for every end-host. Under this condition, the cost term $C^{\ast}$ of the PI equilibrium $\mathbf{F}^+$ to end-hosts can be derived to be $C^{\ast}(\mathbf{F}^+) = d^{p+1}\big(1 - (p/K)/(p/K+1)^{(p+1)/p}\big)$ (cf. Appendix \ref{app:benefits:pi}). The Price of Anarchy to end-hosts under the PI assumption is $$\mathit{PoA}^{\ast+} = \Big(1 - \frac{p/K}{(p/K+1)^{(p+1)/p}}\Big) \cdot \mathit{PoA}^{\ast 0} \leq \mathit{PoA}^{\ast 0}.$$

The cost $C^{\#}$ of the PI equilibrium $\mathbf{F}^+$ to network operators is $C^{\#}(\mathbf{F}^+) = (m + 1/(p/K+1)) \cdot d^p$ and the corresponding Price of Anarchy to network operators is $$\mathit{PoA}^{\#+} = \frac{m+1/(p/K+1)}{m} \leq \frac{m+1}{m} = \mathit{PoA}^{\#0}.$$

Based on the Prices of Anarchy in Table \ref{tab:benefits:poas}, Theorem \ref{theorem:beneficial} holds. However, the Prices of Anarchy $\mathit{PoA}^{\ast+}$ and $\mathit{PoA}^{\#+}$ under the PI assumption are dependent on $K$, which is the number of end-hosts in the network. 
If $K$ is very high, as it is in an Internet context, the Prices of Anarchy under the PI assumption approximate the Prices of Anarchy under the LI assumption. Thus, for scenarios of heterogeneous parallel paths to a destination, the benefit provided by perfect information is undone in an Internet context.
In fact, the effect of additional information can even turn negative when considering more general networks, as we will show in the next section.

\begin{table}
	\vspace{-15pt}
	\renewcommand{\arraystretch}{1.5}
	\centering
	\begin{tabular}{| c | c | c |}
		\cline{2-3}
		\multicolumn{1}{c|}{} & \textbf{LI equilibrium} & \textbf{PI equilibrium}\\
		\hline
		\makecell{\textbf{End-host} \textbf{perspective}} & \(\frac{1}{1 - p/(p+1)^{(p+1)/p}}\)  & \(\frac{1 - (p/K)/(p/K+1)^{(p+1)/p}}{1 - p/(p+1)^{(p+1)/p}}\)\\
		\hline
		\makecell{\textbf{Network-operator} \textbf{perspective}} & \(\frac{m+1}{m}\) & \(\frac{m+1/(p/K+1)}{m}\)\\
		\hline
	\end{tabular}
\vspace{10pt}
	\caption{Price of Anarchy for different perspectives and different equilibrium definitions in the network of parallel links (Figure~\ref{fig:benefits:parallel-links}). \label{tab:benefits:poas}}
	\vspace{-35pt}
\end{table}

\section{The Drawbacks of Information}
\label{sec:harms}

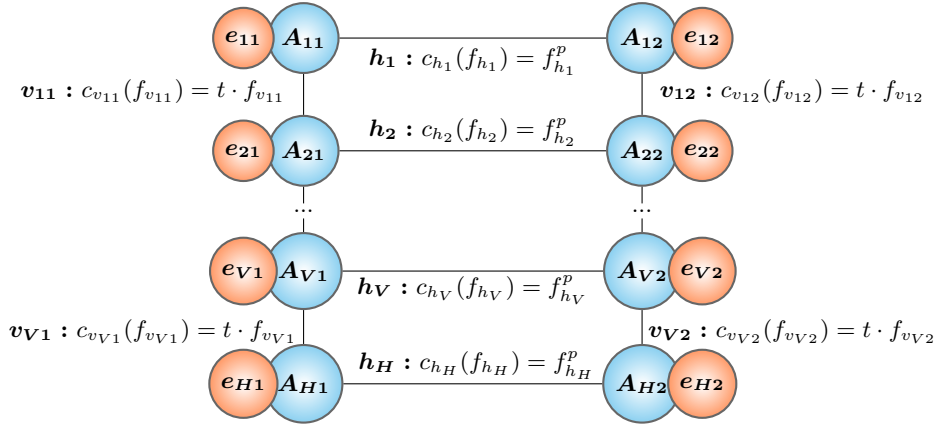
\begin{figure}
	\vspace{-15pt}
	\centering
	\input{fig/network-ladder}
	\vspace{-15pt}
	\caption{Example network illustrating the harmful impact of end-host information (Read: $V = H-1$).}
	\label{fig:harms:ladder-example}
	\vspace{-30pt}
\end{figure}

We will now show that in more general settings, 
more information for end-hosts can \textit{deteriorate} outcomes of selfish routing. Such a case is given by the general \textit{ladder network} in Figure~\ref{fig:harms:ladder-example}, a natural generalization
of the simple topology considered above and a traditional ISP topology \cite{luizelli2013characterizing}.

A ladder network of height $H$ contains $H$ horizontal links $h_1$,..., $h_H$, which represent the rungs of a ladder and have the cost function $c_{h_i}(f_{h_i}) = f_{h_i}^p$. Each horizontal link $h_i$ connects an AS $A_{i1}$ to AS $A_{i2}$, which accommodate the end-hosts $e_{i1}$ and $e_{i2}$, respectively. Every end-host $e_{i1}$ has the same demand $d$ towards the corresponding end-host $e_{i2}$. Neighboring rungs of a ladder are connected by vertical links $v_{ij}$, $i \in \{1,...,V=H-1\}$, $j \in \{1,2\}$, where the vertical link $v_{ij}$ connects the ASes $A_{ij}$ and $A_{i+1,j}$ and has the linear cost function $c_{v_{ij}}(f_{v_{ij}}) = t \cdot f_{v_{ij}}$ with $t \geq 0$. We denote a ladder network with this structure and a choice of parameters $H$, $p$, $d$, and $t$ by~$\mathcal{L}(H,p,d,t)$.

By comparing optima and equilibria, we will prove the following theorem in the following subsections:\begin{theorem}
	For any ladder network $\mathcal{L}(H,p,d,t)$, the Value of Information for both end-hosts and network operators is \textbf{negative}, i.e., $\mathit{VoI}^{\ast} < 0$ and $\mathit{VoI}^{\#} < 0$. \label{thm:ladder}
\end{theorem}

\subsection{Social Optima}
\label{sec:harms:optima}

Both the end-host optimum $\mathbf{F}^{\ast}$ and $\mathbf{F}^{\#}$ are equal to the \textit{direct-only} path-flow pattern $\mathbf{F}^{\sim}$ that is defined as follows: For every end-host $e_{i1}$, $F^{\sim}_{(i1,i2),h_i} = d$  and $F^{\sim}_{(i1,i2),q} = 0$ where $q$ is any other path between $A_{i1}$ and $A_{i2}$ than the direct path over link $h_i$. 

Simple intuition already confirms the optimality of this path-flow pattern. The social cost from the horizontal links is minimized for an equitable distribution of the whole-network demand $Hd$ onto the $H$ horizontal links. In contrast, the cost from vertical links $v_{ij}$ can be minimized to $0$ by simply abstaining from using vertical links. In fact, every use of the vertical links is socially wasteful.

More formally, if $f_{h_i} = d^p$ for $i \in \{1,...,H\}$ and $f_{v_{i1}} = f_{v_{i2}} = 0$ for $i \in \{1,...,V\}$, the marginal costs of the direct path and every indirect path can be
easily shown to equal $(p+1)d^p$, given end-host cost function $C^{\ast}$. 
Concerning network-operator cost $C^{\#}$, the direct and indirect paths have marginal
costs $p\cdot d^{p-1}$ and $p\cdot d^{p-1}+2yt\ \forall y \in \mathbb{N}_{\geq1}$, respectively.
The used direct paths thus do not have a higher marginal cost than the unused indirect paths. 

%
%


\subsection{LI Equilibrium}
\label{sec:harms:li}

Also the LI equilibrium path-flow pattern $\mathbf{F}^0$ is equal to the direct-only path-flow pattern $\mathbf{F}^{\sim}$.
For $\mathbf{F}^{\sim}$, the cost of a direct path $\pi$ is $C_{\pi}(\mathbf{F}^{\sim}) = F_{(i1,i2),\pi}^p = d^p$ and the cost of an indirect path $\pi'$ is $f_{h'}^p + \sum_{v \in W_{\pi'}} f_v = d^p + 0 = d^p$, where $\pi'$ contains the remote horizontal link $h'$ and the vertical links $v \in W_{\pi'}$. Thus, the LI equilibrium conditions of cost equality are satisfied by~$\mathbf{F}^{\sim}$.

As the LI equilibrium is equal to the social optimum both from the end-host perspective and the network-operator perspective, both variants of the Price of Anarchy under the LI assumption are optimal, i.e., $\mathit{PoA}^{\ast0} = \mathit{PoA}^{\#0} = 1$.

\subsection{PI Equilibrium}
\label{sec:harms:pi}

Differently than under the LI assumption, the direct-only flow distribution $\mathbf{F}^{\sim}$ is not stable under the PI assumption. An end-host $e_i$ can improve its individual cost by allocating some traffic to an indirect path $\pi_{k}$ (involving the horizontal link $h_k$) and interfering with another end-host $e_k$. This reallocation decision will increase the social cost for end-hosts and network operators.  In particular, the end-host $e_k$ that previously used the link $h_k$ exclusively will see its selfish cost increase. In turn, the harmed end-host $e_k$ will reallocate some of its traffic to an indirect path in order to reduce its selfish cost $C_{(e_k)}$, leading to a process where all end-hosts in the network interfere with each other until they reach a PI equilibrium with a suboptimal social cost for end-hosts and network operators.

Similar to \S\ref{sec:benefits:pi}, we use the condition of marginal selfish cost equality in order to derive the Price of Anarchy under the PI assumption for a ladder network with $H = 2$. This derivation, as performed in Appendix \ref{app:harms:pi}, yields the following results for the Price of Anarchy to end-hosts and network operators:$$\mathit{PoA}_{H=2}^{\ast+}(p) = 1 + p/12 \quad \quad \mathit{PoA}_{H=2}^{\#+}(p) = 1 + p/3$$

Since the the LI equilibrium is optimal and the PI equilibrium is generally suboptimal
on the considered ladder networks, Theorem \ref{thm:ladder} holds. This finding is confirmed by a case study of the Abilene network (cf. \S\ref{sec:case-study}), which structurally resembles a ladder topology. The case study also reveals that the negative impact of information is
especially pronounced if path diversity is high. 

Interestingly, there is an upper bound of the Price of Anarchy to network operators for a \emph{general} ladder network. This bound is given by the following theorem and proven in Appendix \ref{app:harms:theorem}:

\begin{theorem}
	For every ladder network $\mathcal{L}(H,p,d,t)$, the Price of Anarchy $\mathit{PoA}^{\#+}$ to network operators is lower than the following upper bound $\mathit{PoA}_{\max}^{\#+}$:$$\mathit{PoA}^{\#+} \leq \mathit{PoA}_{H,\max}^{\#+} = 1 + \frac{2(H-1)}{3H}p \leq \mathit{PoA}_{\max}^{\#+} = 1 + \frac{2}{3}p$$
	\label{theorem:ladder-poa}
	\vspace{-10pt}
\end{theorem}

\section{Case Study: Abilene Network}
\label{sec:case-study}

To verify and complement our theoretical insights, we conducted a case study with a real network: we consider the well-known Abilene network, for which topology and workload data is publicly available \cite{knight2011internet, kolaczyk2009statistical}. We accommodate the Abilene topology into our model as follows. For the demand $d$ between the 11 points-of-presence, which we consider ASes, we rely on the empirical traffic matrix from the dataset. Concerning the link-cost functions $c_{\ell}$, we model the latency behavior of a link by a function $c_{\ell}(f_{\ell}) = f_{\ell}^2 + \delta_{\ell}$, where $f_{\ell}^2$ captures the queueing delay and $\delta_{\ell}$ is a constant quantity depending on the geographical distance between the two end-points of link $\ell$, approximating the link's propagation delay.

In order to study the effect of both end-host information and multi-path routing on the Price of Anarchy, we perform the following simulation experiment. First, we compute the social optima $\mathbf{F}^{\ast}$ and $\mathbf{F}^{\#}$ for the Abilene network. Second, we simulate the convergence to the Nash equilibria $\mathbf{F}^{0}$ and $\mathbf{F}^{+}$ for different degrees of multi-path routing, represented by the maximum number of shortest paths that end-hosts consider in their path selection. Once converged, we compute the social cost of the equilibrium traffic distributions and the corresponding Prices of Anarchy. 

\begin{figure}
	\centering
	\includegraphics[width=\linewidth]{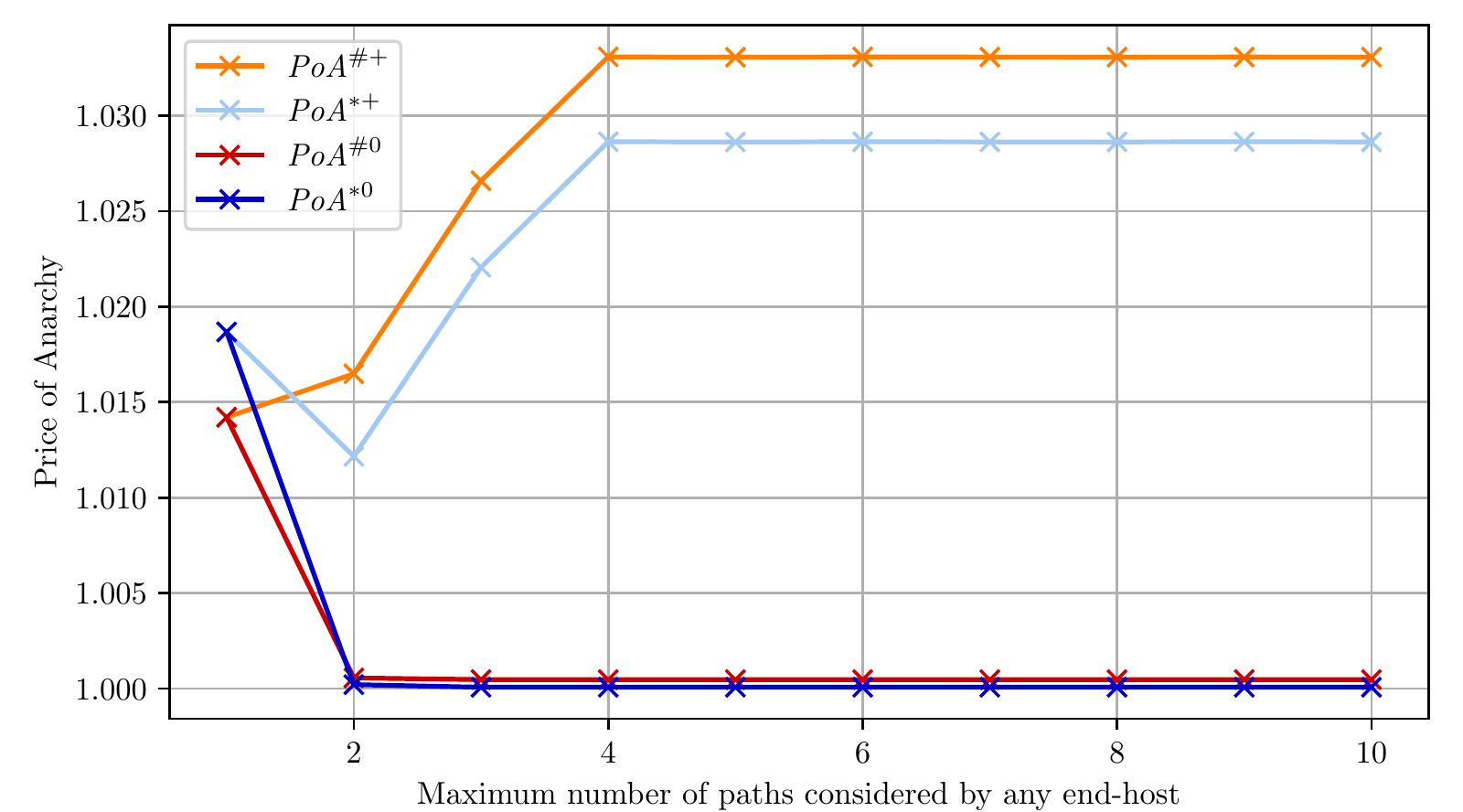}
	\caption{Abilene network results.}
	\label{fig:case-study:abilene-results}
	\vspace{-15pt}
\end{figure}

The experiment results in Figure \ref{fig:case-study:abilene-results} offer multiple interesting insights. Most prominently, if simple shortest-path routing represents the baseline of network-controlled path selection, source-based path selection with latency-only information improves the performance of the network (up to a near-optimum), which confirms findings of prior work \cite{qiu2003selfish}. In contrast, path selection with perfect information deteriorates performance, especially for a higher degree of multi-path routing. Therefore, the potential performance benefits of source-based path selection with multi-path routing are conditional on the amount of information possessed by end-hosts, where a higher degree of information is associated with lower performance. However, while an increasing degree of multi-path routing is associated with worse performance under perfect information, the resulting inefficiency is bounded at a modest level of less than 4 percent for both end-hosts and network operators. The near-optimality of latency-only information in terms of performance and the bounded character of the Price of Anarchy under perfect information reflect the findings from \S\ref{sec:harms} about ladder topologies, which resemble the Abilene topology. Thus, the experiment results not only show that source-based path selection can be a means to improve the performance of a network but also confirm the practical relevance of our theoretical findings.

\section{Related Work}
\label{sec:related-work}

Inefficiency arising from selfish behavior in networks is well-known to exist in transportation networks and has been thoroughly analyzed with the framework of the Wardrop model~\cite{wardrop1952some,dafermos1969traffic}. The most salient expressions of this inefficiency is given by the Braess Paradox~\cite{braess1968paradoxon}.

Literature on \textit{selfish routing} is often concerned with the discrepancy between optimum and Nash equilibrium: the \textit{Price of Anarchy}~\cite{koutsoupias1999worst,dubey1986inefficiency}. The Price of Anarchy was initially studied for network models (see Nisan et al.~\cite{nisan2007algorithmic}
for an overview), but literature now covers a wide spectrum, 
from health care to basketball~\cite{roughgarden2015price}. 
Our work has a closer connection to more traditional research questions, such as
bounds on the Price of Anarchy for selfish routing. An early result has been obtained by Koutsoupias and Papadimitriou~\cite{koutsoupias1999worst}, who formulated routing in a network of parallel links as a multi-agent multi-machine scheduling problem.

A different model has been developed by Roughgarden and Tardos~\cite{roughgarden2002bad} who build on the Wardrop model~\cite{wardrop1952some} for routing in the context of computer networks. The Price of Anarchy in the proposed routing game is the ratio between the latency experienced by all users in the Wardrop equilibrium and the minimum latency experienced by all users. For different classes of latency functions, the authors derive explicit high bounds on the resulting Price of Anarchy. In a different work, they show that the worst-case Price of Anarchy for a function class can always be revealed by a simple network of parallel links and that the upper bound on the Price of Anarchy depends on the growth rate of the latency functions~\cite{roughgarden2003price}.

The relatively loose upper bounds on the Price of Anarchy of previous works \cite{koutsoupias1999worst, roughgarden2002bad} have been qualified by subsequent research. It was found that problem instances with high Prices of Anarchy are usually artificial. By introducing plausible assumptions to make the routing model more realistic, upper bounds on the Price of Anarchy can be reduced substantially. For instance, Friedman~\cite{friedman2004genericity} shows that the Price of Anarchy is lower than the mentioned worst-case derived by Roughgarden and Tardos~\cite{roughgarden2002bad} if the Nash equilibrium cost is not sensitive to changes in the demand of agents. By computing the Price of Anarchy for a variety of different latency functions, topologies, and demand vectors, Qiu et al.~even show that selfish routing is nearly optimal in many cases~\cite{qiu2003selfish}.

Convergence to Nash equilibria has been studied in the context of congestion games \cite{rosenthal1973class} and, in a more abstract form, in the context of potential games \cite{monderer1996potential, sandholm2001potential}. Sandholm \cite{sandholm2001potential} showed that selfish player behavior in potential games leads to convergence to the Nash equilibrium and, under some conditions, even to convergence to the social optimum.
As the question of equilibrium convergence is traditionally studied separately
from the question of equilibrium cost, we do not address convergence issues in this paper.

The study of the effect of incomplete information also has a long
tradition~\cite{harsanyi1967games}, but still poses significant challenges~\cite{roughgarden2015price}. 
Existing literature in this area primarily focuses on
scenarios where players are uncertain about each others' payoffs,
studying alternative notions of equilibria such as
Bayes-Nash equilibria~\cite{singh2004computing}, 
which also leads to alternative definitions of the price
of anarchy such as the 
Bayes-Nash Price of Anarchy~\cite{leme2010pure,roughgarden2015price}
or the price of stochastic anarchy~\cite{chung2008price}.
A common observation of many papers in this area is
that less information can lead to significantly worse equilibria~\cite{roughgarden2015price}.
There is also literature on the impact on the Price of Anarchy
in scenarios where interacting players only have \emph{local} information, 
e.g., the evolutionary price of
anarchy~\cite{opodis19}.

However, much less is known today about the
role of information in games related to \emph{routing}. 
In this context, one line of existing literature is concerned with the recentness of latency information. Most prominently, research on the damage done by stale information in load-balancing problems~\cite{dahlin2000interpreting, mitzenmacher2000useful} has been applied to routing games by Fischer and V\"ocking~\cite{fischer2009adaptive}. This work investigates whether and how rerouting decisions converge onto a Wardrop equilibrium if these rerouting decisions are based on obsolete latency information. Other recent work about the role of information in routing games investigates how the amount of topology information possessed by agents affects the equilibrium cost~\cite{acemoglu2018informational}. 

Existing work on the subject of source-routing efficiency differs from our work in two important aspects. First, to the best of our knowledge, all existing work on the subject defines the social optimum as the traffic assignment that minimizes the total cost experienced by users, which is indeed a reasonable metric. However, our work additionally investigates the total cost experienced by \textit{links}, i.e., the network operators. Since cost considerations by network operators are a decisive factor in the deployment of source-based path selection architectures, the Price of Anarchy to network operators is an essential metric. Second, although existing work on the topic has investigated the role played by the \textit{recentness} of congestion information or the degree of \emph{topology} information, it does not investigate the role played by the \textit{degree of congestion information} that agents possess. Indeed, a major contribution of our work is to highlight the effects of perfect information, i.e., information that allows agents to perfectly minimize their selfish cost. Latency-only information, which agents are assumed to have in existing work, does not enable agents to perform perfect optimization.

\section{Conclusion}
\label{sec:conclusion}

Motivated by emerging path-aware network architectures,
we refined and extended the Wardrop model in order to study
the implications of source-based path selection. Our analysis
provides several interesting insights with practical relevance. First, the cost of selfish routing to network operators differs from the cost experienced by users. Since network operators are central players in the adoption of path-aware networking, research on the effects of selfish routing thus needs to address the network-operator perspective separately. However, we proved upper bounds on the Price of Anarchy which suggest that selfish routing imposes a low cost on network operators. 
Second, we found that basic latency information, which can be measured by the end-hosts themselves, leads to near-optimal traffic allocations in many cases. Selfish routing thus causes modest ineffiency even if end-hosts have only imperfect path information and network operators do not disseminate detailed path-load information.

Our model and first results introduce several
exciting avenues for future research. 
First, we note that while we have focused
on path-aware network architectures, we hope to apply our model
to other practical applications where source routing
has currently received much attention,
e.g., in the context of segment routing~\cite{filsfils2015segment} and multi-cast \cite{shahbaz2018elmo}.
Furthermore, we aim to obtain a more general understanding of the 
interactions between the network topology structure 
and the Price of Anarchy in selfish routing.
Moreover, our focus in this paper was on
rational players, and it is important to extend
our model to account for other behaviors, e.g., players 
combining altruistic, selfish and Byzantine behaviors.

\section{Acknowledgements}
We gratefully acknowledge support from ETH Zurich, from SNSF for
project ESCALATE (200021L\_182005), and from WWTF for project
WHATIF (ICT19-045, 2020-2024). Moreover, we thank Markus Legner, Jonghoon Kwon, and Juan A. García-Pardo for helpful discussions that
supported and improved this research. Lastly, we thank the anonymous
reviewers for their constructive feedback.


%


%

%
\bibliographystyle{splncs04}
\bibliography{bibliography}


\begin{subappendices}
	\renewcommand{\thesection}{\Alph{section}}%

\section{Proofs}
\label{app}

\subsection{Parallel Links: End-Host Optimum}
\label{app:benefits:endhost-optimum}

In the end-host optimum $\mathbf{F}^{\ast}$, it holds that for every end-host $e_k$, $k \in \{1,...,K\}$, the marginal cost of the path over the link $\beta$ is equal to the marginal cost of any path over a link $\alpha_i$, $i \in \{1,...,m\}$. Using this insight, the end-host optimum can be derived by the solution of the following equation: $$\frac{\partial}{\partial F_{(k,D),\beta}} \hspace{5pt} C^{\ast} = \frac{\partial}{\partial F_{(k,D),\alpha_i}} \hspace{5pt} C^{\ast}$$. 

As it holds that $$\frac{\partial}{\partial F_{(k,D),\ell}} C^{\ast} = \frac{\partial}{\partial f_{\ell}} C^{\ast}$$
the optimal link-flow pattern $\mathbf{f}^{\ast}$ is obtained by solving: $$\frac{\partial}{\partial f_{\beta}} (f_{\beta} \cdot f_{\beta}^p) = \frac{\partial}{\partial f_{\alpha_i}} (f_{\alpha_i} \cdot d^p),$$ which yields $f_{\beta}^{\ast} = d/\sqrt[p]{p+1}$.

Any path-flow pattern $\mathbf{F}^{\ast}$ that produces a link-flow pattern with $f_{\beta} = f_{\beta}^{\ast}$ and $\sum_{\alpha_i} f_{\alpha_1} = d - f_{\beta}^{\ast}$ is thus optimal from the perspective of end-hosts. The total cost of such an optimal path-flow pattern $\mathbf{F}^{\ast}$ is given by $$C^{\ast}(\mathbf{F}^{\ast}) = f_{\beta}^{\ast}\cdot {f_{\beta}^{\ast}}^p + (d - f_{\beta}^{\ast}) \cdot d^p = d^{p+1}\Big(1 - p/(p+1)^{(p+1)/p}\Big)$$

\subsection{Parallel Links: PI Equilibrium}
\label{app:benefits:pi}

In the network from Figure \ref{fig:benefits:parallel-links}, the selfish cost function of an end-host $e_k$ can be simplified to $$C^{\ast}_{(e_k)}(\mathbf{F}) = F_{(k,D),\beta} \cdot f_{\beta}^p + \sum_{\alpha_i} F_{(k,D),\alpha_i} \cdot d^p$$. 

Since it holds that $\partial/\partial F_{(k,D),\beta} \hspace{5pt} f_{\beta} = 1$, the marginal selfish costs for the paths over link $\beta$ and $\alpha_i$ are given by \begin{equation*}
\begin{split}
\partial/\partial F_{(k,D),\beta} \hspace{5pt} C^{\ast}_{(e_k)}(\mathbf{F}) &= f_{\beta}^p + F_{(k,D),\beta} \cdot p \cdot f_{\beta}^{p-1}\\
\partial/\partial F_{(k,D),\alpha_i} \hspace{5pt} C^{\ast}_{(e_k)}(\mathbf{F}) &= d^p.
\end{split}
\end{equation*}

An equilibrium under the PI assumption is characterized by the equality of these selfish marginal costs.

Note that for every end-host $e_k$, the marginal selfish cost of every path over a link $\alpha_i$ is the same, namely $d^p$. By marginal selfish cost equality, the marginal selfish costs $\partial/\partial F_{(k,D),\beta} \big(C^{\ast}_{(e_k)}\big)$ must be equal to $d^p$ for every end-host $e_k$ and thus also equal across all end-hosts. This condition is only satisfied if every end-host $e_k$ has the same flow on the path over link $\beta$, i.e., $F_{(m,D),\beta} = F_{(n,D),\beta}$ for all end-hosts $e_m, e_n$. As $f_{\beta} = \sum_{e_k} F_{(k,D), \beta}$, the fact $\sum_{e_k} F_{(k,D), \beta} = K \cdot F_{(k,D), \beta}$ implies $F_{(k,D),\beta} = f_{\beta}/K$ for every end-host $e_k$.

This knowledge about $F_{(k,D),\beta}$ allows to simplify the selfish marginal cost equation and to compute the PI equilibrium. By inserting $f_{\beta}/K$ for $F_{(k,D),\beta}$, the selfish marginal cost equality reads $$f_{\beta}^p + f_\beta/K \cdot p \cdot f_{\beta}^{p-1} = d^p.$$ 

The solution of this equation yields the PI equilibrium link flow $f_{\beta}^+ = d/\sqrt[p]{p/K+1}$. The PI equilibrium $\mathbf{F}^{+}$ is thus given by every path-flow pattern that satisfies the following conditions for every end-host $e_k$: $$F_{(k,D),\beta} = f_{\beta}^+/K \quad \land \quad \sum_{\alpha_i} F_{(k,D),\alpha_i} = (d-f_{\beta}^+)/K.$$

The cost term $C^{\ast}$ of the PI equilibrium $\mathbf{F}^+$ to end-hosts is thus $$C^{\ast}(\mathbf{F}^+) = d^{p+1}\Big(1 - (p/K)/(p/K+1)^{(p+1)/p}\Big).$$

\subsection{Ladder Network: PI Equilibrium}
\label{app:harms:pi}


In order to compute the Price of Anarchy under the PI assumption for a general ladder network, we start by computing the Price of Anarchy for the simple ladder network of $H = 2$. Conforming to the PI equilibrium conditions in \S\ref{sec:model:equilibria:pi}, the PI equilibrium is given by the solution of the following equilibrium equation system $\mathcal{E}_2$ that formalizes the marginal cost equality: $$\begin{cases}
f_{h_1}^p + F_1'\ p\ f_{h_1}^{p-1} = f_{h_2}^p + t (f_{v_{11}} + f_{v_{12}}) + F_1 \big(p \cdot f_{h_2} + 2t\big)\\
f_{h_2}^p + F_2'\ p\ f_{h_2}^{p-1} = f_{h_1}^p + t (f_{v_{11}} + f_{v_{12}}) + F_2 \big(p \cdot f_{h_1} + 2t\big)
\end{cases}$$

\noindent where $F_1' = F_{(11,12),h_1}$ and $F_1 = F_{(11,12),v_{11}h_2v_{12}}$ and the analogous abbreviations have been made by $F_2'$ and $F_2$ for the direct and indirect path flow of end-host $e_{21}$. Due to demand constraints, it holds that $F_1' = d - F_1$ and $F_2' = d - F_2$. Due to the symmetry of the equation system, it is possible to conclude that $F_1' = F_2'$ and $F_1 = F_2$. Since $f_{h_1} = F_1' + F_2$ and $f_{h_2} = F_2' + F_1$, using the symmetry yields $f_{h_1} = f_{h_2} = d - F_1 + F_1 = d$. Furthermore, the flow on the vertical links $v_{11}$ and $v_{12}$ can be expressed as follows: $f_{v_{11}} = f_{v_{22}} = F_1 + F_2 = 2F_1$. The equilibrium equation system can thus be reduced to the single equation: \begin{equation*}
\begin{split}
&d^p + (d-F_1) \cdot \lambda = d^p + 2t\cdot(2F_1) + F_1 \cdot \big(\lambda + 2t\big)\\
\iff&  \lambda d - (6t+2\lambda) F_1 = 0 \iff F_1 = \lambda d/(6t+2\lambda)
\end{split}
\end{equation*} where $\lambda = p\cdot d^{p-1}$.

Based on this solution for the path flow $F_1 = F_{(11,12),v_{11}h_2v_{12}}$, all other path flows can be derived, which yields the following terms for the two perspectives on the Price of Anarchy: \begin{equation*}
\begin{split}
\mathit{PoA}_{H=2}^{\ast+}(d,t,p) &= \big(2d^{p+1} + 2t(2\lambda d/(6t+2\lambda))^2\big)/\big(2\cdot d^{p+1}\big)\\
\mathit{PoA}_{H=2}^{\#+}(d,t,p) &= \big(2d^p + 4t\lambda d/(6t+2\lambda)\big)/\big(2\cdot d^p\big)
\end{split}
\end{equation*}

The Price of Anarchy for all ladder networks with $H=2$ is obtained by computing an upper bound on the Price of Anarchy in terms of demand $d$ and parameter $t$: $$\mathit{PoA}_{H=2}^{\ast+}(p) = 1 + p/12 \hspace{15pt} \mathit{PoA}_{H=2,\max}^{\#+}(p) = 1 + p/3$$

\subsection{Ladder Network: Proof of Theorem \ref{theorem:ladder-poa}}
\label{app:harms:theorem}

We start by observing that $\mathit{PoA}_{H,\max}^{\#+}$ is given by the limit in~$t$ and is only dependent on the flows on vertical links $v$: 
$$\mathit{PoA}_{H,\max}^{\#+} = 1 + \lim_{t \rightarrow \infty} (t\cdot \sum_{v} f_v)/(H\cdot d^p),$$ 
where we used that  $\lim_{t \rightarrow \infty} \sum_{h} f_h^p = H\cdot d^p$ as vertical links become infinitely expensive. We need only characterize the sum of vertical link flows $f_V = \sum_{v} f_v$, for which we use an argument based on the structure of the equilibrium equation system $\mathcal{E}_H$. 

\begin{figure}
	\centering
	\includegraphics[width=\columnwidth]{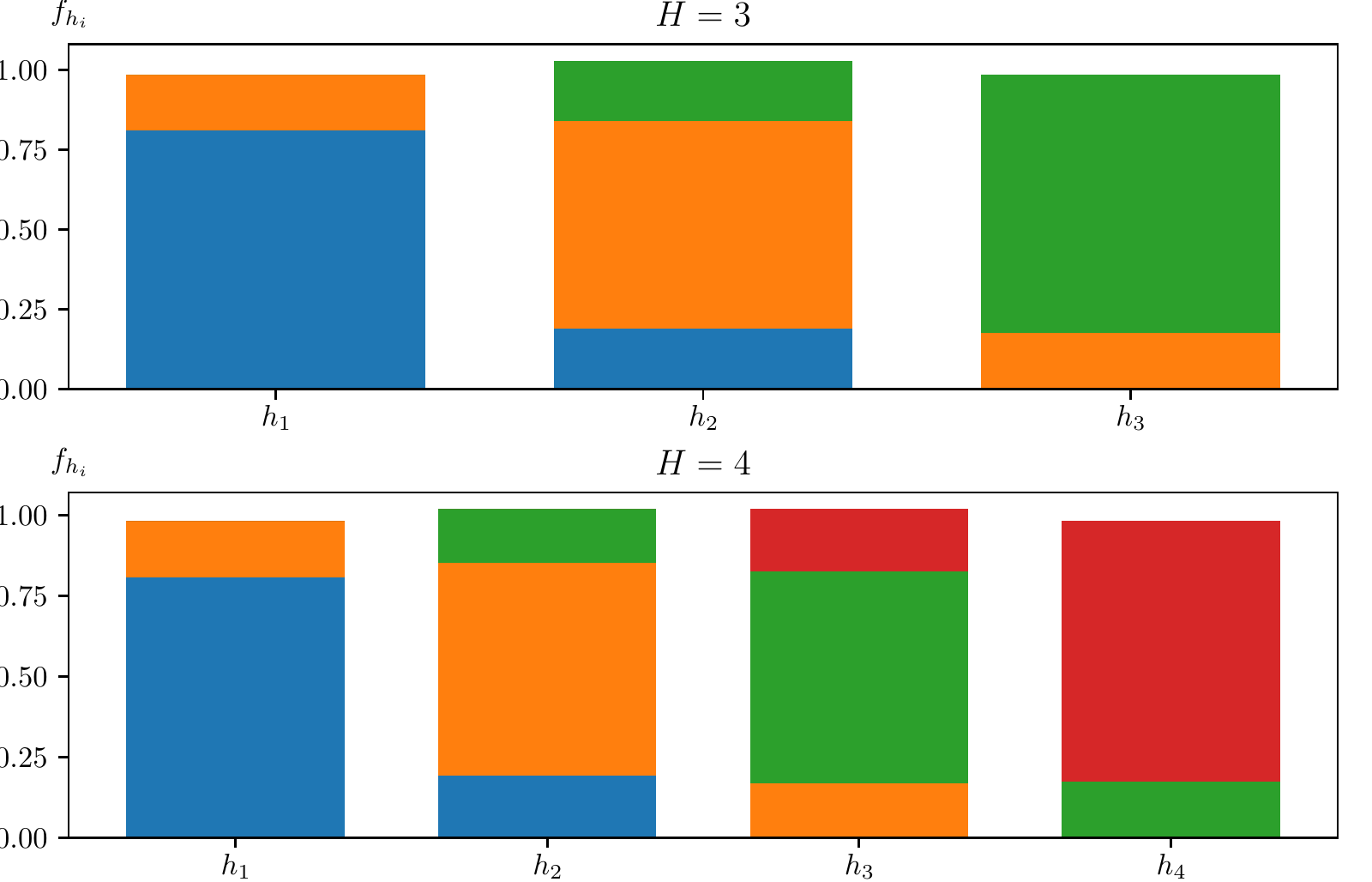}
	\caption{Traffic distribution over horizontal links of a ladder network in PI equilibrium $\mathbf{F}^{+}$ (for $p =2$, $t = 1$, $H = 3,4$).}
	\label{fig:harms:ladder-traffic-distribution}
	\vspace{-10pt}
\end{figure}

\begin{figure}
	\centering
	\begin{subfigure}{0.4\columnwidth}
		\input{fig/ladder-variables-H3}
		\caption{H = 3}
		\label{fig:harms:ladder-variables:3}
	\end{subfigure}
	\begin{subfigure}{0.4\columnwidth}
		\input{fig/ladder-variables-H4}
		\caption{H = 4}
		\label{fig:harms:ladder-variables:4}
	\end{subfigure}
	\caption{Variable assignments to indirect path flows in a ladder network for computing the PI equilibrium.}
	\label{fig:harms:ladder-variables}
	\vspace{-10pt}
\end{figure}
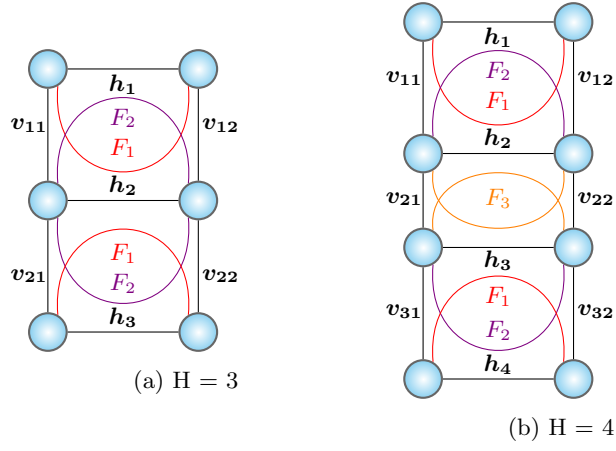

For setting up $\mathcal{E}_H$, we consider Figure~\ref{fig:harms:ladder-traffic-distribution} which illustrates numerically computed equilibria for some $H > 2$. The figure shows the traffic distribution on the horizontal links of a ladder network, where different-color flow shares correspond to flows of different end-hosts. Figure~\ref{fig:harms:ladder-traffic-distribution} shows two insights that are relevant for setting up $\mathcal{E}_H$. First, the path-flow pattern $\mathbf{F}^{+}$ only contains non-zero flows on paths that deviate at most one ladder level from the originating end-host (for high enough $t$). Second, the path-flow pattern is symmetric with respect to the horizontal axis of the ladder network. The variables in $\mathcal{E}_H$ can thus be assigned to the indirect path flows as displayed in Figure~\ref{fig:harms:ladder-variables}. Variable assignments for higher $H$ work analogously to Figure~\ref{fig:harms:ladder-variables:3} (for odd $H$) and Figure~\ref{fig:harms:ladder-variables:4} (for even $H$). 

With these variables and the knowledge about the equilibrium traffic distribution, the equation system $\mathcal{E}_H$ can be set up for all values for $H$. Table~\ref{tab:harms:ladder-equations} lists the equation systems $\mathcal{E}_H$ for all $H$. All equations in a system $\mathcal{E}_H$ are of the form $E_{a,b}(F_{c}, F_{d}, F_{f}) = 0$, where $$E_{a,b}(F_{c}, F_{d}, F_{f}) = \lambda d - (at+b\lambda)\cdot F_c - 2t\cdot F_d - \lambda \cdot F_f.$$ The set $E(\mathcal{E}_H)$ contains all left-hand side terms $E_{a,b}$ of the equations in $\mathcal{E}_H$. Let the sum $\Sigma(\mathcal{E}_H)$ of an equation system $\mathcal{E}_H$ be the equation $\sum_{E(\mathcal{E}_H)} E_{a,b} = 0$. It holds that for all $H$, $\Sigma(\mathcal{E}_H)$ is the equation $$(H-1)\lambda d - (6t+2\lambda)\cdot F_1 - (6t + 3\lambda) \sum_{2\leq u \leq H-1} F_u = 0.$$ Solving equation $\Sigma(\mathcal{E}_H)$ for $F_V = \sum_{1 \leq u \leq H-1} F_u$, we obtain $$F_V = \frac{(H-1)\lambda d}{\rho(t)\cdot(6t+3\lambda)}$$ where $\lim_{t \rightarrow \infty} \rho(t) = 1$.

\begin{table}
	\renewcommand{\arraystretch}{1.5}
	\centering
	\begin{tabular}{| c | c |}
		\hline
		$\boldsymbol{H}$ & $\boldsymbol{\mathcal{E}_H}$\\
		\hline
		$H = 2$ & \(\begin{cases}E_{6,2}(F_1, 0, 0) = 0\end{cases}\)\\
		\hline
		$H = 3$ & \(\begin{cases}
		E_{4,2}(F_1, F_2, 0) = 0\\
		E_{4,3}(F_2, F_1, 0) = 0\\
		\end{cases}\)\\
		\hline
		\makecell{Even $H \geq 4$\\Odd $j$, $3\leq j \leq H-3$} & \(\begin{cases}
		E_{4,2}(F_1, F_2, 0) = 0\\
		E_{4,2}(F_2, F_1, F_3) = 0 \vspace{5pt}\\
		E_{4,2}(F_j, F_{j+1}, F_{j-1}) = 0\\
		E_{4,2}(F_{j+1}, F_{j}, F_{j+2}) = 0 \vspace{5pt}\\
		E_{6,2}(F_{H-1}, 0, F_{H-2}) = 0
		\end{cases}\)\\
		\hline
		\makecell{Odd $H \geq 5$\\Odd $j$, $3 \leq j \leq H-4$} & \(\begin{cases}
		E_{4,2}(F_1, F_2, 0) = 0\\
		E_{4,2}(F_2, F_1, F_3) = 0 \vspace{5pt}\\
		E_{4,2}(F_j, F_{j+1}, F_{j-1}) = 0\\
		E_{4,2}(F_{j+1}, F_{j}, F_{j+2}) = 0 \vspace{5pt}\\
		E_{4,2}(F_{H-2}, F_{H-1}, F_{H-3}) = 0\\
		E_{4,3}(F_{H-1}, F_{H-2}, 0) = 0
		\end{cases}\)\\
		\hline
	\end{tabular}
	\caption{Equation systems $\mathcal{E}_H$ characterizing PI equilibrium for all $H$. \label{tab:harms:ladder-equations}}
	\vspace{-10pt}
\end{table}

Due to the horizontal and vertical symmetry of the PI equilibrium on the ladder network, it holds that $f_V = 4 \cdot F_V$. Inserting $f_V$ into $\mathit{PoA}_{H,\max}^{\#+}$ yields $$1 + \lim_{t \rightarrow \infty} \frac{t}{H\cdot d^p} \cdot \frac{4(H-1)\lambda d}{\rho(t)(6t+3\lambda)}= 1+ \frac{2(H-1)}{3H}\cdot p.$$ 

Taking the limit of this term for $H \rightarrow \infty$ results in $$\mathit{PoA}^{\#+}_{\max} = 1 + 2/3 \cdot p.$$

\end{subappendices}

\end{document}

%% file: comment-macros.tex
\newif\ifremovecomments
\removecommentsfalse

\makeatletter
\newcommand{\mysubscript}[2]{\textsubscript{\textcolor{#2}{\textsf{\textbf{#1}}}}}
\newcommand*\defcomment[4]{
  \ifremovecomments
    \expandafter\newcommand\csname #1\endcsname[1]{%
    }
    \expandafter\newcommand\csname @#2delnoname\endcsname[1]{%
    }
    \expandafter\newcommand\csname #2del\endcsname[1]{%
    }
    \expandafter\newcommand\csname #2sugg\endcsname[1]{##1}
    \expandafter\newcommand\csname #2subs\endcsname[2]{##2}
  \else
    \expandafter\newcommand\csname #1\endcsname[1]{%
        \textcolor{#4}{\ding{110}\mysubscript{#3}{#4}\,{##1}}%
    }
    \expandafter\newcommand\csname @#2delnoname\endcsname[1]{%
        \bgroup\markoverwith{\textcolor{#4}{\rule[0.35ex]{2pt}{1pt}}}\ULon{##1}%
    }
    \expandafter\newcommand\csname #2del\endcsname[1]{%
        \csname @#2delnoname\endcsname{##1}\kern0.1em\mysubscript{#3}{#4}%
    }
    \expandafter\newcommand\csname #2sugg\endcsname[1]{%
        \textcolor{#4}{[##1]\mysubscript{#3}{#4}}%
    }
    \expandafter\newcommand\csname #2subs\endcsname[2]{%
        \csname @#2delnoname\endcsname{##1}\csname #2sugg\endcsname{##2}%
    }
  \fi
    \expandafter\newcommand\csname #2sout\endcsname{\csname #2del\endcsname}
}
\makeatother
\defcomment{simon}{si}{SI}{blue}
\defcomment{stefan}{st}{ST}{magenta}

%% file: fig/model-example.tex
\begin{tikzpicture}[
	asnode/.style={circle, draw=black!60, shading=radial,outer color={rgb,255:red,137;green,207;blue,240},inner color=white, thick, minimum size=7mm},
	agentnode/.style={circle, draw=black!60, shading=radial,outer color={rgb,255:red,255;green,153;blue,102},inner color=white, thick, minimum size=4mm},
	]
	
	\node[asnode]	 (A1)	  at (0.7, 2)	{$\boldsymbol{A_1}$};
	\node[asnode]	 (A3)	  at (6.3, 2)	{$\boldsymbol{A_3}$};
	\node[asnode]	 (A2)	  at (3.5, 3.5)	{$\boldsymbol{A_2}$};
	
	\draw[-] (A1.north east) -- (A2.south west);
	\node[] at (0.8, 2.8) {$\boldsymbol{\alpha:}$ $c_{\alpha}(f_{\alpha}) = 1$};
	\draw[-] (A2.south east) -- (A3.north west);
	\node[] at (6.6, 2.89) {$\boldsymbol{\beta:}$ $c_{\beta}(f_{\beta}) = f_{\beta}^2$};
	\draw[-] (A1.east)       -- (A3.west);
	\node[] at (4.5, 2.2) {$\boldsymbol{\gamma:}$ $c_{\gamma}(f_{\gamma}) = f_{\gamma}$};
	
	\node[] (start1) at (0, 2.25) {};
	\node[] (start11) at (0, 1.75) {};
	\node[] (end1)   at (5, 3.75) {};
	\draw[-,blue!70!black, line width=1mm] (start1) to[spline through={(1.4,2.5)(3.3,3)}] (4.2,3.5);
	\node[blue!70!black] at (3.7,2.65) {$\big[(1,4),\alpha\big]$};
	\draw[-,violet, line width=1mm] (start11) to[spline through={(1.5,1.7)(3.5,1.8)(6,2)}] (4.2,3.5);
	\node[violet] at (2.2,2.2) {$\big[(1,4),\gamma\beta\big]$};
	\node[] (start2)   at (2.0, 3.75) {};
	\node[] (end2)     at (7.0, 1.75) {};
	
	\node[agentnode] (e1)     at (0,   2)	{$\boldsymbol{e_1}$};
	\node[agentnode] (e2)     at (2.8,   3.5)	{$\boldsymbol{e_2}$};
	\node[agentnode] (e3)     at (7.0, 2)	{$\boldsymbol{e_3}$};
	\node[agentnode] (e4)     at (4.2,   3.5)	{$\boldsymbol{e_4}$};

\end{tikzpicture}

%% file: fig/introduction-example.tex
\begin{tikzpicture}[
	asnode/.style={circle, draw=black!60, shading=radial,outer color={rgb,255:red,137;green,207;blue,240},inner color=white, thick, minimum size=7mm},
	agentnode/.style={circle, draw=black!60, shading=radial,outer color={rgb,255:red,255;green,153;blue,102},inner color=white, thick, minimum size=4mm},
	]
	\node[asnode]	 (asO)	  at (1.5,1.5)	{\textbf{O}}; 
	\node[asnode]	 (asD)	  at (7.5,1.5)	{\textbf{D}};
	\node[agentnode] (agent1) at (0,  1.5)	{\textbf{e}};
	\node at (4.5,2.3) {$\boldsymbol{\alpha}$: $c_{\alpha}(f_{\alpha}) = f_{\alpha} + \frac{1}{2}$};
	\node at (4.5,0.7) {$\boldsymbol{\beta}$: $c_{\beta}(f_{\beta}) = 2$};
	
	\draw[-] (agent1.east) -- (asO.west);
	\draw[-] (asO.north east) to [out=10,in=170] (asD.north west);
	\draw[-] (asO.south east) to [out=350,in=190](asD.south west);
	
	\draw[-,color=blue!70!black] (asO.35) to [out=10,in=170] (asD.145);
	\node[blue!70!black] at (4.5,1.75) {$f_{\alpha} - F_{\alpha} = 1$};
	
	\draw[-,color=red!70!black] (asO.325) to [out=350,in=190](asD.215);
	\node[red!70!black] at (4.5,1.25) {$f_{\beta} - F_{\beta} = 1$};
\end{tikzpicture}

%% file: fig/network-parallel-links.tex
\begin{tikzpicture}[
	asnode/.style={circle, draw=black!60, shading=radial,outer color={rgb,255:red,137;green,207;blue,240},inner color=white, thick, minimum size=7mm},
	agentnode/.style={circle, draw=black!60, shading=radial,outer color={rgb,255:red,255;green,153;blue,102},inner color=white, thick, minimum size=4mm},
	]
	\node[asnode]	 (asO)	  at (1.5,1.5)	{\textbf{O}}; 
	\node[asnode]	 (asD)	  at (7.5,1.5)	{\textbf{D}};
	\node[agentnode] (agenta) at (1,2)	{$\boldsymbol{e_1}$};
	\node[agentnode] (agentb) at (1,1)	{$\boldsymbol{e_K}$};
	\node at (4.5,2.3) {$\boldsymbol{\alpha_1}$: $c_{\alpha_1}(f_{\alpha_1}) = d^p$};
	\node at (4.5,1.5) {$\boldsymbol{\alpha_m}$: $c_{\alpha_m}(f_{\alpha_m}) = d^p$};
	\node at (4.5,1.9) {...};
	\node at (4.5,0.7) {$\boldsymbol{\beta}$: $c_{\beta}(f_{\beta}) = f_{\beta}^p$};
	\node at (0.95,1.5) {...};
	
	\draw[-] (asO.north east) to [out=10,in=170] (asD.north west);
	\draw[-] (asO.east) to [out=10,in=170] (asD.west);
	\draw[-] (asO.south east) to [out=350,in=190](asD.south west);
\end{tikzpicture}

%% file: fig/network-ladder.tex
\begin{tikzpicture}[
	asnode/.style={circle, draw=black!60, shading=radial,outer color={rgb,255:red,137;green,207;blue,240},inner color=white, thick, minimum size=7mm},
	agentnode/.style={circle, draw=black!60, shading=radial,outer color={rgb,255:red,255;green,153;blue,102},inner color=white, thick, minimum size=4mm},
	]
	\node[asnode]	 (asA11)	  at (1.5, 2.3)	{$\boldsymbol{A_{11}}$}; 
	\node[asnode]	 (asA12)	  at (6, 2.3)	{$\boldsymbol{A_{12}}$};
	\node[asnode]	 (asA21)	  at (1.5, 0.8)	{$\boldsymbol{A_{21}}$}; 
	\node[asnode]	 (asA22)	  at (6, 0.8)	{$\boldsymbol{A_{22}}$};
	\node[asnode]	 (asA31)	  at (1.5, -0.8)	{$\boldsymbol{A_{V1}}$}; 
	\node[asnode]	 (asA32)	  at (6, -0.8)	{$\boldsymbol{A_{V2}}$};
	\node[asnode]	 (asA41)	  at (1.5, -2.3)	{$\boldsymbol{A_{H1}}$}; 
	\node[asnode]	 (asA42)	  at (6, -2.3)	{$\boldsymbol{A_{H2}}$};
	\node[agentnode] (e11) at (0.7,2.3)	{$\boldsymbol{e_{11}}$};
	\node[agentnode] (e21) at (0.7,0.8)	{$\boldsymbol{e_{21}}$};
	\node[agentnode] (e31) at (0.7,-0.8)	{$\boldsymbol{e_{V1}}$};
	\node[agentnode] (e41) at (0.7,-2.3)	{$\boldsymbol{e_{H1}}$};
	\node[agentnode] (e12) at (6.8,2.3)	{$\boldsymbol{e_{12}}$};
	\node[agentnode] (e22) at (6.8,0.8)	{$\boldsymbol{e_{22}}$};
	\node[agentnode] (e32) at (6.8,-0.8)	{$\boldsymbol{e_{V2}}$};
	\node[agentnode] (e42) at (6.8,-2.3)	{$\boldsymbol{e_{H2}}$};
	
	\draw[-] (asA11.east)  to (asA12.west);
	\node at (3.75, 2) {$\boldsymbol{h_1:}$ $c_{h_1}(f_{h_1}) = f_{h_1}^p$};
	\draw[-] (asA21.east)  to (asA22.west);
	\node at (3.75, 1.05) {$\boldsymbol{h_2:}$ $c_{h_2}(f_{h_2}) = f_{h_2}^p$};
	\draw[-] (asA11.south) to (asA21.north);
	\node at (-0.5, 1.6) {$\boldsymbol{v_{11}:}$ $c_{v_{11}}(f_{v_{11}}) = t \cdot f_{v_{11}}$};
	\draw[-] (asA12.south) to (asA22.north);
	\node at (8, 1.6) {$\boldsymbol{v_{12}:}$ $c_{v_{12}}(f_{v_{12}}) = t \cdot f_{v_{12}}$};
	
	\draw[-] (asA21.south) to (1.5, 0.1);
	\draw[-] (asA22.south) to (6, 0.1);
	\node at (1.5, 0) {...};
	\node at (6, 0) {...};
	\draw[-] (asA31.north) to (1.5, -0.1);
	\draw[-] (asA32.north) to (6, -0.1);
	
	\draw[-] (asA31.east)  to (asA32.west);
	\node at (3.75, -1.05) {$\boldsymbol{h_{V}:}$ $c_{h_{V}}(f_{h_{V}}) = f_{h_{V}}^p$};
	\draw[-] (asA41.east)  to (asA42.west);
	\node at (3.8, -2.05) {$\boldsymbol{h_H:}$ $c_{h_H}(f_{h_H}) = f_{h_H}^p$};
	\draw[-] (asA31.south) to (asA41.north);
	\node at (-0.5, -1.6) {$\boldsymbol{v_{V1}:}$ $c_{v_{V1}}(f_{v_{V1}}) = t \cdot f_{v_{V1}}$};
	\draw[-] (asA32.south) to (asA42.north);
	\node at (8, -1.6) {$\boldsymbol{v_{V2}:}$ $c_{v_{V2}}(f_{v_{V2}}) = t \cdot f_{v_{V2}}$};
	
	\end{tikzpicture}

%% file: fig/ladder-variables-H3.tex
\begin{tikzpicture}[
		asnode/.style={circle, draw=black!60, shading=radial,outer color={rgb,255:red,137;green,207;blue,240},inner color=white, thick, minimum size=5mm},
		]
		
		\node[asnode]	 (asA11)	  at (0, 1.75)	{}; 
		\node[asnode]	 (asA12)	  at (2, 1.75)	{};
		\node[asnode]	 (asA21)	  at (0, 0)	{}; 
		\node[asnode]	 (asA22)	  at (2, 0)	{};
		\node[asnode]	 (asA31)	  at (0, -1.75)	{}; 
		\node[asnode]	 (asA32)	  at (2, -1.75)	{};
		
		\draw[-] (asA11.east)  to (asA12.west);
		\node[]at (1, 1.55) {$\boldsymbol{h_1}$};
		\draw[-,color=red] (asA11.300) .. controls (asA21) and (asA22) .. (asA12.240);
		\node[color=red] at (1, 0.7) {$F_{1}$};
		
		\draw[-] (asA21.east)  to (asA22.west);
		\node[]at (1, 0.2) {$\boldsymbol{h_2}$};
		\draw[-,color=violet] (asA21.60)  .. controls (asA11) and (asA12) .. (asA22.120);
		\node[color=violet] at (1, 1.1) {$F_{2}$};
		\draw[-,color=violet] (asA21.300) .. controls (asA31) and (asA32).. (asA22.240);
		\node[color=violet] at (1, -1.1) {$F_{2}$};
		
		\draw[-] (asA31.east)  to (asA32.west);
		\node[]at (1, -1.55) {$\boldsymbol{h_3}$};
		\draw[-,color=red] (asA31.60)  .. controls (asA21) and (asA22).. (asA32.120);
		\node[color=red] at (1, -0.7) {$F_{1}$};
		
		\draw[-] (asA11.south) to (asA21.north);
		\node[] at (-0.25, 1) {$\boldsymbol{v_{11}}$}; 
		\draw[-] (asA12.south) to (asA22.north);
		\node[] at (2.3, 1) {$\boldsymbol{v_{12}}$}; 
		\draw[-] (asA21.south) to (asA31.north);
		\node[] at (-0.25, -1) {$\boldsymbol{v_{21}}$}; 
		\draw[-] (asA22.south) to (asA32.north);
		\node[] at (2.3, -1) {$\boldsymbol{v_{22}}$}; 
		
		\end{tikzpicture}

%% file: fig/ladder-variables-H4.tex
\begin{tikzpicture}[
		asnode/.style={circle, draw=black!60, shading=radial,outer color={rgb,255:red,137;green,207;blue,240},inner color=white, thick, minimum size=5mm},
		]
		
		\node[asnode]	 (asA11)	  at (0, 1.75)	{}; 
		\node[asnode]	 (asA12)	  at (2, 1.75)	{};
		\node[asnode]	 (asA21)	  at (0, 0)	{}; 
		\node[asnode]	 (asA22)	  at (2, 0)	{};
		\node[asnode]	 (asA31)	  at (0, -1.25)	{}; 
		\node[asnode]	 (asA32)	  at (2, -1.25)	{};
		\node[asnode]	 (asA41)	  at (0, -3)	{}; 
		\node[asnode]	 (asA42)	  at (2, -3)	{};
		
		\draw[-] (asA11.east)  to (asA12.west);
		\node[]at (1, 1.55) {$\boldsymbol{h_1}$};
		\draw[-,color=red] (asA11.300) .. controls (asA21) and (asA22) .. (asA12.240);
		\node[color=red] at (1, 0.7) {$F_{1}$};
		
		\draw[-] (asA21.east)  to (asA22.west);
		\node[]at (1, 0.2) {$\boldsymbol{h_2}$};
		\draw[-,color=violet] (asA21.60)  .. controls (asA11) and (asA12) .. (asA22.120);
		\node[color=violet] at (1, 1.1) {$F_{2}$};
		\draw[-,color=orange] (asA21.300) .. controls (asA31) and (asA32).. (asA22.240);
		\node[color=orange] at (1, -0.6) {$F_{3}$};
		
		\draw[-] (asA31.east)  to (asA32.west);
		\node[]at (1, -1.42) {$\boldsymbol{h_3}$};
		\draw[-,color=orange] (asA31.60)  .. controls (asA21) and (asA22) .. (asA32.120);
		\draw[-,color=violet] (asA31.300) .. controls (asA41) and (asA42).. (asA32.240);
		\node[color=violet] at (1, -2.35) {$F_{2}$};
		
		\draw[-] (asA41.east)  to (asA42.west);
		\node[]at (1, -2.8) {$\boldsymbol{h_4}$};
		\draw[-,color=red] (asA41.60)  .. controls (asA31) and (asA32).. (asA42.120);
		\node[color=red] at (1, -1.9) {$F_{1}$};
		
		\draw[-] (asA11.south) to (asA21.north);
		\node[] at (-0.25, 1) {$\boldsymbol{v_{11}}$}; 
		\draw[-] (asA12.south) to (asA22.north);
		\node[] at (2.3, 1) {$\boldsymbol{v_{12}}$}; 
		\draw[-] (asA21.south) to (asA31.north);
		\node[] at (-0.25, -0.6) {$\boldsymbol{v_{21}}$}; 
		\draw[-] (asA22.south) to (asA32.north);
		\node[] at (2.3, -0.6) {$\boldsymbol{v_{22}}$};
		\draw[-] (asA31.south) to (asA41.north);
		\node[] at (-0.25, -2.1) {$\boldsymbol{v_{31}}$}; 
		\draw[-] (asA32.south) to (asA42.north);
		\node[] at (2.3, -2.1) {$\boldsymbol{v_{32}}$}; 
		
		\end{tikzpicture}